\documentclass
[lengthcheck,10pt,tightenlines,twocolumn,superscriptaddress,pre,nofootinbib]{revtex4}%
\usepackage{amsfonts}
\usepackage{amsmath}
\usepackage{amssymb}
\usepackage{graphicx}%
\providecommand{\U}[1]{\protect\rule{.1in}{.1in}}
\newtheorem{theorem}{Theorem}
\newtheorem{acknowledgement}[theorem]{Acknowledgement}

\begin{document}
\title{Entropy of complex relevant components of Boolean networks }
\author{Peter Krawitz}
\affiliation{Institute for Systems Biology, Seattle, WA 98103, USA }
\affiliation{Fakult\"{a}t f\"{u}r Physik, Ludwig Maximilians Universit\"{a}t, 80799
M\"{u}nchen, Germany}
\author{Ilya Shmulevich}
\affiliation{Institute for Systems Biology, Seattle, WA 98103, USA }

\begin{abstract}
Boolean network models of strongly connected modules are capable of capturing
the high regulatory complexity of many biological gene regulatory circuits. We
study numerically the previously introduced basin entropy, a parameter for the
dynamical uncertainty or information storage capacity of a network as well as
the average transient time in random relevant components as a function of
their connectivity. We also demonstrate that basin entropy can be estimated
from time-series data and is therefore also applicable to non-deterministic
networks models.

\end{abstract}
\maketitle

\section{Introduction}

Random Boolean networks are often studied as generic models of gene regulatory
networks \cite{KauffmanJTB69}\cite{KauffmanPNAS03}. The ensemble approach to
gene regulation, a term coined by Kauffman, aims at studying well-defined
ensembles of model networks, the statistical features of which match those of
real cells and organisms \cite{KauffmanPA04}. Ensembles of special biological
interest are critical random Boolean networks, which lie at the boundary
between a frozen and a chaotic phase \cite{DerridaEPL86}\cite{DerridaEPL86b}%
\cite{ShmulevichPNAS05}. In the frozen phase, a perturbation to one node
propagates on average to less than one other node during one time step. In the
chaotic phase, the difference between two almost identical states increases
exponentially fast, since a perturbation propagates on average to more than
one node during one time step \cite{Aldana03}. Critical Boolean networks
balance robustness against perturbations with complex asymptotic attractor dynamics.

Since Boolean networks are deterministic systems, they eventually settle into
periodic attractors. Regarding the behavior in this asymptotic limit, nodes
can be classified into three groups. Nodes that are frozen to the same value
on every attractor form the frozen core\textit{ }of a network
\cite{FlyvbjergJPA88b}. These nodes give a constant input to other nodes and
are otherwise irrelevant. Nodes that are not frozen but have no outputs, or
only outputs to other irrelevant nodes, are also classified as irrelevant.
Although they might fluctuate, they do not influence the number and periods of
attractors. Finally, the relevant nodes are those non-frozen nodes that
influence their own state over directed loops. The recognition of the relevant
nodes as the only elements influencing the asymptotic dynamics was an
important step in understanding the dynamics of Boolean networks
\cite{BastollaPD98a}\cite{BastollaPD98b}.

In a biological context, an attractor is associated with a characteristic
dynamically stable pattern of gene expression, which may represent a
particular fate of the cell. The weight of each such attractor can be defined
as the probability for a random state in the state space of the network to
flow into this attractor. Based on the state space partition into attractors
of different weights, we recently introduced a new network parameter, called
the basin entropy (hereafter, simply entropy), which measures the uncertainty
of the future behavior of a random state. This entropy was shown to increase
with system size in critical network ensembles, whereas for ensembles in the
ordered phase and in highly chaotic networks, it approaches a finite value
\cite{Krawitz07}. From an informational processing perspective, this means
that the complexity of a network increases only with its system size if it
operates near the critical regime.

An intuitive understanding of this growing complexity are networks whose
relevant nodes are modularly organized and whose complexity increases if new
modules accumulate. In living systems, transcriptional regulation is often
highly modular \cite{ShenNat02}\cite{LeeScie02}. Of special interest are
complex relevant components, which consist of relevant nodes containing more
than one relevant input/output \cite{KaufmanNJP06}. Boolean network models for
several biological gene regulatory circuits have been constructed and were
shown to reproduce experimentally observed results \cite{AlbertJTB03}%
\cite{SotoPlant04}\cite{FaureBio06}\cite{LiPNAS04}. These often
highly-connected subnetworks can be viewed as biological examples of such
complex relevant components.

In this work, we first numerically study the entropy of complex relevant
components as a function of their connectivity. We show that the probability
of such a component to freeze increases with growing connectivity and that its
entropy decreases. Additionally, we also study the average transient time of a
random state until it falls into its attractor. The calculation of dynamic
network parameters, such as the basin entropy, requires the knowledge of the
underlying network functions. This often limits the applicability of such
parameters. We demonstrate that the entropy of a network can also be estimated
from time series data by the application of clustering techniques. This
broadens the applicability of this dynamic network parameter to models whose
functions are not necessarily known or that are not deterministic. In order to
illustrate our results, we will use a Boolean network model for the mammalian
cell cycle as presented in \cite{FaureBio06}.

\section{Boolean networks}

In a Boolean network every gene is identified by a node $i$, whose state
$x_{i}\in\left\{  0,1\right\}  $ indicates whether the gene is switched on or
off. Each node $i$ receives input from $k_{i}$ other nodes and its state at
the next time step $t+1$ is a Boolean function $f_{i}$ of the states of the
nodes it is depending on: $x_{i}(t+1)=f_{i}(x_{i_{1}}\left(  t\right)
,...,x_{i_{k_{i}}}\left(  t\right)  )$. A network is entirely defined by its
directed connectivity graph $G$ and the Boolean functions $\mathbf{F}%
=\{f_{1},...,f_{n}\}$ assigned to every node. The state of a network
$\mathbf{x}\left(  t\right)  =\left(  x_{1}\left(  t\right)  ,...,x_{n}\left(
t\right)  \right)  $ contains the values of all $n$ nodes in the network at a
given time point $t$. A Boolean network thus defines a deterministic
transition matrix $\mathbf{T}$ on its state space. A random state
$\mathbf{x}\in\left\{  0,1\right\}  ^{n}$ that is propagated through the
network as%
\begin{equation}
\mathbf{x}(t+1)=\mathbf{F}(\mathbf{x}(t))
\end{equation}
generates a time series or trajectory through the state space that finally
intersects with itself. The states that are then revisited infinitely often
define the attractor $\rho$, with the number of states on the attractor equal
to $l(\rho),$ also called the attractor size. Transient states that lead into
an attractor form its basin of attraction. The sum of all attractor states and
basin states of a certain attractor normalized by the size of the entire state
space ($2^{n}$) define the weight $w_{\rho}$ of that attractor. The weight of
an attractor is the probability that a randomly chosen state will flow into
this very attractor. The entropy $h$ of the probability distribution defined
by $w_{\rho}$ is called the \textit{basin entropy:}%

\begin{equation}
h=-\sum_{\rho}w_{\rho}\ln w_{\rho}.
\end{equation}
This measure captures the uncertainty of the future dynamic behavior of the
system started in a random state. The scaling of this parameter relative to
system size was discussed in \cite{Krawitz07} for ensembles of different
dynamical regimes.

The sensitivity of a node $i$ specifies the number of its input arguments
that, when toggled, will result in a flip in the value of that node, averaged
over all input combinations. A Boolean function $f_{i}$ with $k_{i}$ input
variables that takes on the value $1$ for any one of its possible $2^{k_{i}}$
input vectors with probability $p_{i}$ has the expected sensitivity
\cite{ShmulevichPRL04}:%
\begin{equation}
s_{i}=2k_{i}p_{i}(1-p_{i}) \label{sensitivity}%
\end{equation}

The network sensitivity $s$ is the average sensitivity of all its nodes and
indicates to how many nodes a perturbation to a single node is, on average,
propagated. The network sensitivity is an order parameter of a network
ensemble that divides random Boolean networks with an expected sensitivity of
$s<1$ into the ordered phase and with $s>1$ into the chaotic phase. Random
networks that propagate a perturbation, on average, to one other node ($s=1$),
are called critical. Classical Kauffman networks with a fixed indegree $k=2$
and $p=0.5$, are prototypes of critical Boolean network ensembles, though it
should be noted that this definition of criticality is independent of the
actual indegree distribution.

The attractor dynamics of a Boolean network are entirely determined by its
relevant nodes. The scaling of these nodes was first discussed in
\cite{SocolarPRL03} and derived for a general class of critical network
ensembles by Drossel, Kaufman and Mihaljev \cite{KaufmanPRE05}%
\cite{MihaljevPRE06}.They presented a stochastic process that removes frozen
nodes and nonfrozen irrelevant nodes from a network and ends when there are
only potentially relevant nodes left. We call these nodes `potentially
relevant', as some of them may be part of self-freezing loops. We will discuss
this phenomenon below in more detail. The number of these potentially relevant
nodes $n_{r}$ scales in all critical networks of size $n$ and average
connectivity $k>1$ as $n_{r}\sim n^{1/3}$. In the limit of large $n$, almost
all these relevant nodes depend on only one relevant node; the
\textit{proportion} of relevant nodes depending on more than one relevant
input approaches zero; and the \textit{number} of nodes depending on more than
two relevant inputs is almost surely zero. These results agree with structural
findings of random graphs at the point of phase transition, where the number
of nodes in complex components scales with $n^{1/3}$ and each such complex
component is almost surely topologically equivalent to a 3-regular multigraph
\cite{LuczakTAMS94}.

\section{Random complex relevant components}

A set of relevant nodes that eventually influence each other's state form a
relevant component. We define the excess $e$ of the connection digraph of a
component in analogy to the graph theoretic terminology as the difference
between the number of arrows $a$ between the nodes and the number of nodes
$n$:%
\begin{equation}
e=a-n.
\end{equation}

The topology of the simplest relevant component is a loop of nodes. The excess
of this component is $e=0$. If we randomly add a new link to this loop we have
an intersected loop of the same size with excess $e=1$. One node now depends
on two nodes and one node influences two nodes \footnote{In the Boolean
network literature only relevant components with an excess $e>0$ are called
complex components. This might cause confusion for readers familiar with the
graph theoretic terminology, where a component is called complex if its excess
is $e\geq0$.}.

\begin{figure}
[ptb]
\begin{center}
\includegraphics[
height=1.0473in,
width=3.371in
]%
{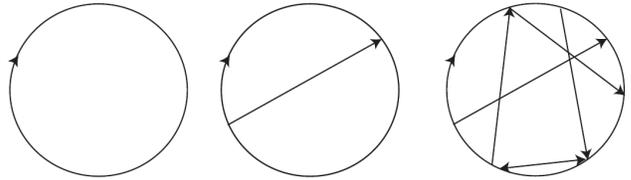}%
\caption{Random complex component of excess $e=0,1,6$ from left to right.}%
\label{fig1}%
\end{center}
\end{figure}

By randomly adding further edges, we may generate components of arbitrary
excess (see Fig. \ref{fig1}).%

In a simple relevant loop only the Boolean `copy' ( $x_{i}(t+1)=x_{j}(t)$ )
and `negation' ( $x_{i}(t+1)=\overline{x_{j}}(t)$ ) functions can be used. A
perturbation in a loop node is always propagated to a single successor node
and therefore such components have sensitivity $s=1$. If another edge is
added, the Boolean function of the node receiving input from two relevant
inputs has to be changed. Generally, whenever the indegree of a node increases
from $k$ to $k+1$ variables, a new Boolean updating function for $k+1$
variables has to be assigned. By adjusting the bias $p$ (see eq.
(\ref{sensitivity})), we can randomly generate a Boolean function of $k+1$
variables so that its expected sensitivity is $E\left(  s\right)  =1$. If any
of the $k+1$ variables happen to be fictitious (i.e., toggling their value has
no effect on the output), we can draw again in order to guarantee that all
added edges are irreducible. This process thus generates random components of
arbitrary excess, operating in the critical or slightly supracritical regime.

\section{Entropy of relevant components}

The only relevant component whose entropy we can discuss analytically is the
simple loop. In simple loops all states are attractor states and it is
therefore sufficient to know the length $l(\rho)$ of an attractor in order to
determine its weight, $w_{\rho}=l(\rho)/2^{n}$. Regarding the attractor
dynamics, we can substitute an even number of `negation' functions in a loop
by `copy' functions, so that it is sufficient to discuss loops with an even
number or an odd number of `negations'. In even loops, the attractor states
that only differ by a rotation of their values belong to the same attractor
and attractors can therefore be viewed as an equivalence class under rotation.
In combinatorics such an equivalence class is also known as a binary necklace
of length $n$. The number of attractors of a simple even loop is calculated as
follows \cite{deBruijn46}:%

\begin{equation}
N_{A}^{even}(n)=\frac{1}{n}\sum_{d|n}\phi\left(  \frac{n}{d}\right)  2^{d}%
\end{equation}
where $\phi(m)$ is Euler's totient function, which is defined as the number of
positive integers $\leq m$ that are relatively prime to $m$ (i.e., do not
contain any factor in common with $m$). The sum is taken over all dividers of
$n$. If $n$ is prime, the number of attractors of length $n$ is simply
$\left(  2^{n}-2\right)  /n$.

In odd loops, a state $\mathbf{x}$ and its negation $\overline{\mathbf{x}}$
are always part of the same attractor and the number of attractors can be
calculated with the formula:%

\begin{equation}
N_{A}^{odd}(n)=\left\{
\begin{array}
[c]{ll}%
\frac{1}{2n}\sum_{d|n}\phi(n/d)2^{d}+\frac{3}{4}2^{\frac{n}{2}}, & \text{if
}n\text{ is even}\\
\frac{1}{2n}\sum_{d|n}\phi(n/d)2^{d}+2^{\frac{n-1}{2}}, & \text{if }n\text{ is
odd}%
\end{array}
\right.
\end{equation}
For $n$ prime, the number of attractors of length $2n$ is $\left(
2^{n}-1\right)  /2n$. For large even (odd) loops, most attractors are of
length $n$ ($2n$) and the entropy can be approximated by considering only
those attractors that are of maximal weight, $h(n)\approx n\ln2-O(\ln n)$.
Therefore, the entropy of simple loops scales linearly with its size under
synchronous updating.

In terms of the entropy, the key difference between complex components with
excess $e\geq1$ compared to simple loops is that the attractor length and
weight are no longer correlated. In complex components, attractors of length
one may have even higher weights than larger attractors in the same network.
The mean number and length of attractors of a random component are
insufficient in describing its dynamic complexity and we choose to study the
entropy as a function of increasing excess\footnote{For a thorough and mainly
analytical discussion of the mean number and length of attractors in relevant
components of excess $e=1,$ see \cite{KaufmanEPJB05}}.

\begin{figure}
[ptb]
\begin{center}
\includegraphics[
height=2.6749in,
width=3.371in
]%
{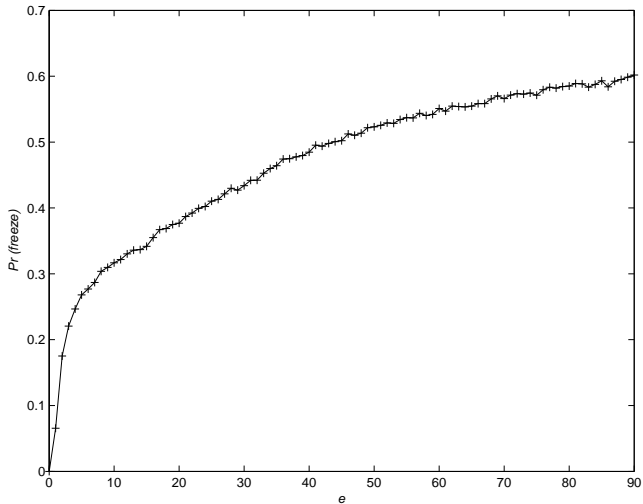}%
\caption{The probability of a $n=10$ node component to freeze increases
roughly logarithmically with increasing excess $e$. }%
\label{fig2}%
\end{center}
\end{figure}

When we start increasing the excess of our component from $e=0$, two
qualitatively different things may happen: either all nodes stay relevant and
only the attractor dynamics change, or parts of the component or perhaps even
the entire component freezes. This special case of self-freezing loops was
first discussed in \cite{PaulPRE06}. The simplest case of a
\textit{self-freezing} component are two nodes that canalize each other to
their majority bits, e.g. $f_{1}=x_{1}\vee x_{2}$, $f_{2}=x_{1}\vee
\overline{x_{2}}$. Such components are clearly not relevant components, but
part of the frozen core of a network. In Fig. \ref{fig2} the probability that
a critical component of $n=10$ nodes freezes is shown for increasing excess
$e$. The average was taken over more than $26,000$ network instances for every
excess $e$. We obtained the same qualitative progress for component sizes up
to $n=18$ \footnote{The computational time to calculate the entropy increases
exponentially with system size. We therefore chose larger increments for the
excess in larger components: $n=12,14,16,18$, $e=1,2,...,10,15,20,...,90$.
Over $2000$ network instances have been simulated for every ensemble.}: The
addition of the first few edges to a loop strongly increases the probability
to freeze, whereas the probability to freeze increases slower in components of
already high excess.%

\begin{figure}
[ptb]
\begin{center}
\includegraphics[
trim=0.339669in 0.000000in 0.656609in 0.235698in,
height=2.8798in,
width=3.3702in
]%
{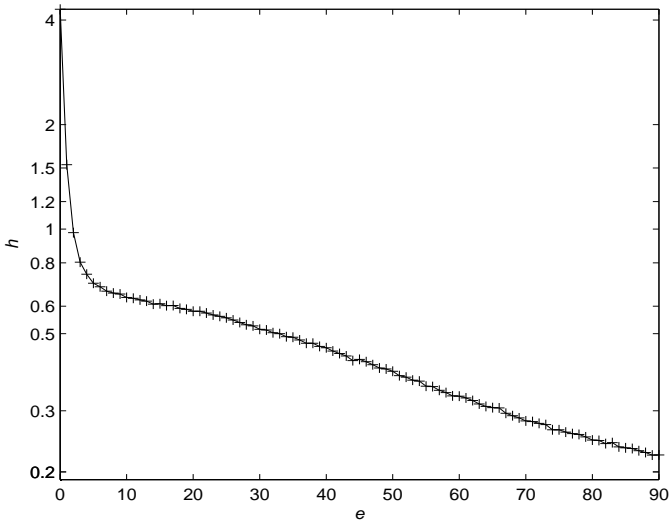}%
\caption{Average entropy $\left\langle h\right\rangle $ of a complex relevant
component with increasing excess $e.$}%
\label{fig3}%
\end{center}
\end{figure}

If a component becomes frozen, or only has a single attractor, it has entropy
$h=0$ and we will not consider it as a relevant component. Fig.\ref{fig3}
shows the decline in the average entropy $\left\langle h\right\rangle $ of the
relevant $n=10$ node components as a function of their excess. The entropy
drops sharply for the first few additional edges and decreases slower for
$e>10$. For all studied component sizes $n$ up to $18$, the average entropy
falls below $1$ for $e=10$ and continues to decrease slower thereafter.%

\begin{figure}
[ptb]
\begin{center}
\includegraphics[
height=2.6809in,
width=3.371in
]%
{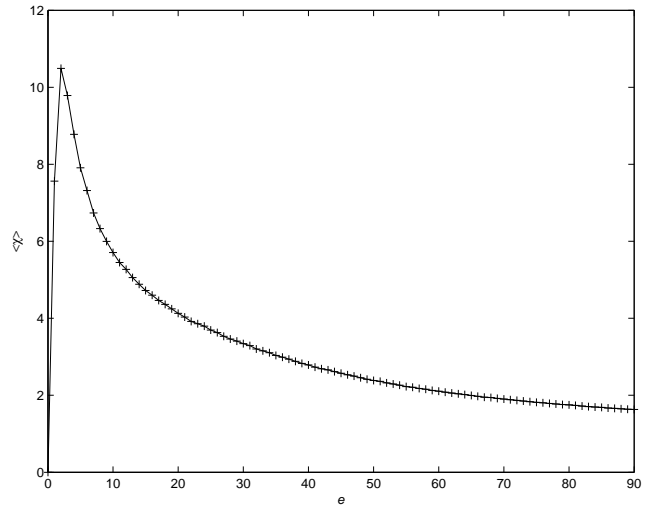}%
\caption{The average transient time decreases after a peak around $e\approx2$
with increasing excess.}%
\label{fig4}%
\end{center}
\end{figure}

As soon as additional edges are introduced into the loop, the average number
of attractor states drops substantially and the majority of states become
transient states. The average number of steps that are required to reach an
attractor from a random state in the state space is defined as the average
transient time $\left\langle \chi\right\rangle $ of a network. In a simple
loop, where all states are attractor states, the transient time is zero. If we
increase the excess, the transient time first sharply increases, peaks around
an excess of $e=2$, $\left\langle \chi\right\rangle _{n=10}^{\max}(e=3)=10$
and begins to decrease thereafter as shown in Fig. \ref{fig4}. We obtain
qualitatively the same behavior for the transient time in components of sizes
up to $n=18$, ($\left\langle \chi\right\rangle _{n=18}^{\max}(e=3)\approx28$,
$\left\langle \chi\right\rangle _{n=18}(e=100)\approx3$).%

\begin{figure}
[ptb]
\begin{center}
\includegraphics[
height=2.6507in,
width=3.371in
]%
{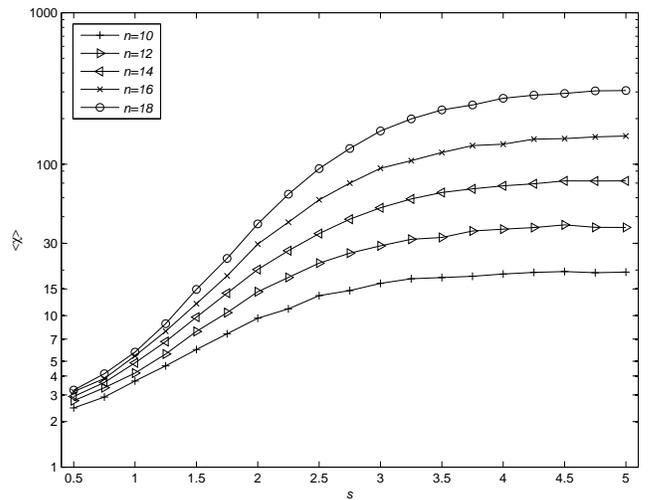}%
\caption{Average transient time $\left\langle \chi\right\rangle $ in random
network ensembles of increasing sensitivity $s$.}%
\label{fig5}%
\end{center}
\end{figure}

We also studied the average transient time as a function of the average
sensitivity (Fig. \ref{fig5}). The average transient time starts to grow
rapidly, soon after the ensembles enter the chaotic regime ($s>1$) and scales
with the system size in highly chaotic networks.%

Compared to the average transient times in networks of higher sensitivity $s$,
even the maximal transient times in the critical relevant components (Fig.
\ref{fig4}) are small. The updating functions of nodes of the critical
component that depend on more than one relevant input are more likely to be
\textit{canalizing}\footnote{A function is canalizing for the value
$\sigma_{i}=\left\{  0,1\right\}  $ of variable $i$ if this value can
determine the function value regardless of the values of the other input
variables.} because of the stronger bias $p$ that was used in their generation
process \cite{Just04}. Canalizing functions are found frequently as control
rules governing the transcription in eukaryotic genes \cite{HarrisComp02}.
Dynamic properties of random Boolean networks with canalizing functions have
also been studied in \cite{KauffmanPNAS03},\cite{PaulPRE06}. A higher
proportion of canalizing functions leads characteristically to short
attractors and more robust dynamics.

\section{Boolean networks modeling biological circuits}

As a showcase model for a biological gene regulatory circuit, we now analyze
the Boolean network of the mammalian cell cycle as presented in
\cite{FaureBio06}. This $n=10$ node network simulates the states of cell cycle
genes that regulate the process of cell division and its quiescent G$_{0}$
phase. It can be viewed as a biological example of a complex relevant
component of highly connected nodes (for a more thorough biological
discussion, see \cite{FaureBio06}). The Boolean functions and attractor states
of this network are shown in the tables below. The value on the left hand side
of the equations corresponds to the value at time $t+1$, $x_{i}%
(t+1)=f(x_{i_{1}}(t),...,x_{i_{k_{i}}}(t)),$ with the symbols $+$ and $\cdot$
representing the OR and AND operations, respectively:%
\[%
\begin{tabular}
[c]{|c|c|}\hline
Gene & Boolean function\\\hline
CycD & \multicolumn{1}{|l|}{$x_{1}=x_{1}$}\\\hline
Rb & \multicolumn{1}{|l|}{$x_{2}=(\overline{x_{4}}\overline{x_{5}}%
+x_{6})\overline{x_{1}}\overline{x_{10}}$}\\\hline
E2F & \multicolumn{1}{|l|}{$x_{3}=(\overline{x_{5}}+x_{6})\overline{x_{2}%
}\overline{x_{10}}$}\\\hline
CycE & \multicolumn{1}{|l|}{$x_{4}=x_{3}\overline{x_{2}}$}\\\hline
CycA & \multicolumn{1}{|l|}{$x_{5}=(x_{3}+x_{5})\left(  \overline{x_{5}}%
+x_{8}\right)  \overline{x_{2}}\overline{x_{7}}$}\\\hline
p27 & \multicolumn{1}{|l|}{$x_{6}=\left(  \overline{x_{4}}\overline{x_{5}%
}+x_{6}\overline{x_{4}}+x_{6}\overline{x_{5}}\right)  \overline{x_{1}%
}\overline{x_{10}}$}\\\hline
Cdc20 & \multicolumn{1}{|l|}{$x_{7}=x_{10}$}\\\hline
Cdh1 & \multicolumn{1}{|l|}{$x_{8}=\overline{x_{5}}\overline{x_{10}}%
+x_{7}+x_{6}\overline{x_{10}}$}\\\hline
UbcH10 & \multicolumn{1}{|l|}{$x_{9}=\overline{x_{8}}+x_{8}x_{9}(x_{7}%
+x_{6}+x_{10})$}\\\hline
CycB & \multicolumn{1}{|l|}{$x_{10}=\overline{x_{7}}\overline{x_{8}}$}\\\hline
\end{tabular}
\]%
\[%
\begin{tabular}
[c]{|c||c|c|c|c|c|c|c|}\hline
\multicolumn{8}{|c|}{Attractors}\\\hline
G$_{0}$ & \multicolumn{7}{|c|}{Cell Cycle}\\\hline
0 & 1 & 1 & 1 & 1 & 1 & 1 & 1\\\hline
1 & 0 & 0 & 0 & 0 & 0 & 0 & 0\\\hline
0 & 0 & 0 & 0 & 0 & 1 & 1 & 1\\\hline
0 & 0 & 0 & 0 & 1 & 0 & 1 & 1\\\hline
0 & 0 & 1 & 1 & 1 & 0 & 0 & 1\\\hline
1 & 0 & 0 & 0 & 0 & 0 & 0 & 0\\\hline
0 & 1 & 0 & 1 & 0 & 0 & 0 & 0\\\hline
1 & 1 & 0 & 0 & 0 & 1 & 1 & 1\\\hline
0 & 1 & 1 & 1 & 0 & 1 & 0 & 0\\\hline
0 & 0 & 1 & 1 & 0 & 0 & 0 & 0\\\hline
\end{tabular}
\]
The attractor of length seven represents the different steps of the cell cycle
phases, G$_{1},$S$,$G$_{2}$ and M, whereas the fixed point attractor
represents the G$_{0}$ phase. Both attractors have the same weight $w=0.5$,
which yields an entropy of $h=\ln2\approx0.69$. If we average over the
sensitivities of the single nodes, we obtain a network sensitivity of $s=1.04$
which lies in the range of the expected average sensitivity for a relevant
component of a critical network. The average transient time of this network is
$\left\langle \chi\right\rangle =4.8$, which is on the order of a random
relevant component with the same excess ($e=24$).

So far, other detailed deterministic Boolean models have only been presented
for a few other gene regulatory circuits. A prominent example among those is a
model for the segment polarity gene network, which governs the embryonic
pattern formation during certain stages of the developmental process in the
fruit fly \textit{Drosophila melanogaster} \cite{AlbertJTB03}. This Boolean
network models the interaction between eleven genes and its products and
defines certain fixed-point (single-state) attractors that can be identified
as stable gene expression patterns in wild-type embryos. For this model, we
also find the dynamic network parameters to lie within the range of critical
complex components: for the network sensitivity we obtain $s=1.02$, the
entropy is $h=2.17$ and the transient time $\left\langle \chi\right\rangle
=3.6$.

Generally, the identification of a single deterministic logical function for a
gene is often difficult for experimental reasons \cite{SotoPlant04}, or
determinism might not even be a desired feature of the modeling approach
itself. For example, probabilistic Boolean networks consider more than just
one Boolean function as possible updating rules for a gene
\cite{ShmulevichBio02}. Also, asynchronous updating schemes lead to
non-deterministic dynamics \cite{GreilPRL05}. We therefore conclude with a
section in which we sketch an approach that enables us to extend the concept
of basin entropy to non-deterministic models.

\section{Entropy estimated from time-series data}

In an unperturbed Boolean network a trajectory that started from a random
point in the state space will finally be caught in a single attractor. If we
allow small random perturbations in the updating procedure, the trajectory
will jump out of its attractor from time to time and may settle in another
attractor. The deterministic dynamics of the unperturbed network give way to a
stochastic (and ergodic) Markov process with transition probabilities
$p_{ij}=P\left(  \mathbf{x}_{t}=j|\mathbf{x}_{t-1}=i\right)  $, such that
$\sum_{j=1}^{2^{n}}p_{ij}=1$ \cite{pbn2}. The transition probabilities can be
arranged in an ordered fashion in a stochastic state transition matrix,%

\begin{equation}
\mathbf{P}=\left(
\begin{array}
[c]{ccc}%
p_{11} & p_{12} & \cdots\\
p_{21} & p_{22} & \cdots\\
\vdots & \vdots & \ddots
\end{array}
\right)  \text{.}%
\end{equation}
Let $\mathbf{\pi}\left(  0\right)  $ be the vector of initial state
probabilities at time $t=0$. We may calculate the state probability
distribution $\mathbf{\pi}\left(  m\right)  $ after $m$ steps:%

\begin{equation}
\mathbf{\pi}(m)=\mathbf{\pi}(0)\mathbf{P}^{m}\text{.}%
\end{equation}
We may further sum up the probabilities of states, leading to the same
attractor, to get a probability distribution $\mathbf{\upsilon}_{m}$ for the
attractors after $m$ time steps. The steady-state probabilities ($m\rightarrow
\infty$) for attractors in Boolean networks with perturbations were studied in
\cite{brun}.

Let us consider the following two-node network defined by the Boolean
functions $x_{1}(t+1)=x_{1}(t)+x_{2}(t)$ and $x_{2}(t+1)=x_{1}(t)\cdot
\overline{x_{2}}(t)+\overline{x_{1}}(t)\cdot x_{2}(t)$ to illustrate the
difference between the weight distribution of the deterministic case and the
attractor probability distribution in the perturbed case.\ When we use a
perturbation probability of $q=\Pr(x_{i}\rightarrow\overline{x_{i}})=0.1$ for
switching the value of a node \textit{after} calculating the successor of a
state, the deterministic transition matrix%

\[
\mathbf{T}=\left(
\begin{array}
[c]{cccc}%
\ 1\  & \ 0\  & \ 0\  & \ 0\ \\
0 & 0 & 0 & 1\\
0 & 0 & 0 & 1\\
0 & 0 & 1 & 0
\end{array}
\right)
\]
is replaced by the stochastic transition matrix%

\[
\mathbf{P}=\left(
\begin{array}
[c]{cccc}%
0.81 & 0.09 & 0.09 & 0.01\\
0.01 & 0.09 & 0.09 & 0.81\\
0.01 & 0.09 & 0.09 & 0.81\\
0.09 & 0.01 & 0.81 & 0.09
\end{array}
\right)  \text{,}%
\]
where the states along the rows and columns are arranged in the usual
lexicographic order (00, 01, 10, 11).

This Boolean network divides its state space into two different\ basins of
attraction: the first one consists of its fixed point attractor $(00)$ while
the second one contains the transient state $(01)$ that is flowing into the
attractor $\left(  10\right)  \rightleftarrows\left(  11\right)  $. Thus, the
weight distribution is $w_{1}=0.25$ and $w_{2}=0.75$. If we solve the steady
state equation $\mathbf{\pi}=\mathbf{\pi P}$, we obtain $\pi_{\left(
00\right)  }=0.201$, $\pi_{\left(  01\right)  }=0.0598$, $\pi_{\left(
10\right)  }=0.3618$, $\pi_{\left(  11\right)  }=0.3775$. We may sum up the
probabilities of states contributing to the same attractor basin to get a
probability distribution $\mathbf{\nu}$ of the basins, $v_{1}=0.201$ and
$v_{2}=0.799$. If the history of a trajectory is unknown, this distribution
describes the probability of the network operating in a certain basin. It can
be estimated with arbitrary precision by sampling the states of a time series.
The perturbation probability $q$ clearly affects $\mathbf{\nu}$; for large
$q$, the network rules (connections, updating functions) lose their importance
and the time series become random.

\begin{figure}
[ptb]
\begin{center}
\includegraphics[
trim=0.993682in 0.000000in 0.996246in 0.000000in,
height=2.0064in,
width=3.3399in
]%
{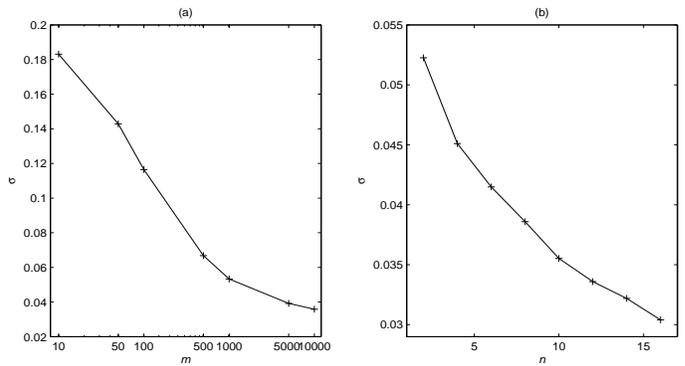}%
\caption{(a): The deviation $\sigma$ between the weight distribution
$\mathbf{w}$ and the estimated basin probability distribution $\mathbf{\nu}$
decreases for longer time series $m$ ($n=10$ and $q=0.01$ fixed). (b): The
deviation $\sigma$ becomes smaller with increasing network size $n$
($m=10,000$ and $q=0.01$ fixed).}%
\label{fig6}%
\end{center}
\end{figure}

For our described two-node example network, the basin weight distribution
$\mathbf{w}$ and the basin probability distribution $\mathbf{\nu}$ differ. We
therefore studied how increasing network size affects the average deviation,%

\begin{equation}
\sigma=\frac{1}{|\rho|}\sqrt{\Sigma_{\rho}\left(  w_{\rho}-\nu_{\rho}\right)
^{2}}%
\end{equation}
between the basin probability distribution $\mathbf{\nu}$ and the weight
distribution $\mathbf{w}$ in different network ensembles. In Fig.
\ref{fig6}(a) the behavior of $\sigma$ is shown when increasingly long time
series with $q=0.01$ are used to estimate the basin probability distribution
of critical $n=10$ network ensembles. Fig. \ref{fig6}(b) shows that the mean
deviation $\sigma$ decreases for increasing network sizes ($q=0.01$ and
$m=10,000$). Therefore, especially in larger networks, the basin weight
distribution may be well estimated by the basin probability distribution. For
all network ensembles the average was taken over more than $10,000$ network instances.%

The assignment of a state in the time series to its attractor, if the
underlying network rules are unknown, is a challenging classification problem.
Two states occurring in the time series with comparable frequency may either
belong to the same attractor or to two different attractors that just happen
to have a similar weight or probability. In order to solve this problem, we
have to make a second assumption: states that belong to the same attractor are
more likely to occur in temporal proximity. Instead of looking at a single
state, we consider all states in a time window of a certain length $\tau$. The
size of the time window $\tau$ generally has to be estimated from the time
series data \cite{Kantz04}. In a perturbed trajectory, generated from a
Boolean network, $\tau$ should be on the order of the expected attractor
lengths. In a model based on a biological circuit, the choice for the expected
length may also be motivated by `biological intuition'.

The classification of these time profiles into several attractors is a
clustering problem, where the number of clusters is not known. Many algorithms
to estimate the optimal number of clusters in a data set have been developed
and extensively studied. Generally, more free parameters (clusters) enable to
further minimize the error function on which the cluster algorithm is based.
This better `fit' is then penalized by a term growing with the number of free
parameters. Based on this trade-off criterion, an `optimal' clustering model
can be found. When dealing with biological data, a range for the number of
expected clusters (e.g. different cell states) can also be provided by the
experimentalist. It is not our intention here to discuss the challenges of
clustering and we refer the interested reader to the extensive literature in
this field \cite{Duda73},\cite{Schwarz78},\cite{jain}. However, for
illustrational purposes, we sketch the analysis of the already introduced
network of the mammalian cell cycle by a perturbed time series and clustering.

We generated a time series of $m=10,000$ successor states from the Boolean
model of the mammalian cell cycle. The value of every node was flipped with
probability $q=0.01$ after calculating the successor state. Profiles were
generated by adding up the values of a node during the time window $\tau$:%
\begin{equation}
c_{i}\left(  t\right)  =\sum_{t^{\prime}=t-\tau}^{t}x_{i}\left(  t^{\prime
}\right)
\end{equation}
with $c_{i}\in\left\{  0,...,\tau+1\right\}  $. Different values of $\tau$
have been tested: $\tau=4,\ldots,10$. The distance between two profiles
$\mathbf{c}=\left(  c_{1},\ldots,c_{n}\right)  $ and $\mathbf{c}^{\prime
}=\left(  c_{1}^{\prime},\ldots,c_{n}^{\prime}\right)  $ was measured by the
city block ($L_{1}$) distance:%
\begin{equation}
d\left(  \mathbf{c,c}^{\prime}\right)  =\sum_{i=1}^{n}\left\vert c_{i}%
-c_{i}^{\prime}\right\vert .
\end{equation}

The profiles were then clustered by the \textit{k-}means algorithm
\cite{Duda73}. In order to determine the optimal number of clusters, the
Bayesian information criterion (BIC) and Akaike information criterion (AIC)
have been used and yielded an optimal number of two clusters for all used
$\tau$. This correctly corresponds to the two attractors of the underlying
Boolean network, the fixed-point attractor of the G$_{0}$ phase and the
attractor of length seven of the cell cycle. The classification of the time
series into two attractors yields a probability distribution $\mathbf{\upsilon
}$, whose entropy $h=0.691$ is close to the `true' network's entropy,
$h\approx\ln2$. This example demonstrates how the attractors and their weight
distribution, a dynamical property of the network, can be derived from a time
series using a straightforward clustering approach that does not require
knowledge about the underlying network rules.

\section{Discussion}

We studied the average entropy and transient time of random complex relevant
components near criticality as a function of their excess. This was motivated
in part by new analytical results on the degree distribution of relevant nodes
in critical network ensembles \cite{MihaljevPRE06}. In random graphs of such
given degree distributions, most nodes are organized with high probability as
a single giant component \cite{CooperCPC04},\cite{MolloyCPC98}. The regulatory
elements in gene networks, on the other hand, seem to be organized in a more
modular manner \cite{ShenNat02},\cite{LeeScie02}. This raises the question of
whether (ordinary) critical random Boolean networks that have been
successfully used as models for the study of gene regulatory dynamics still
lack characteristic properties of their biological archetypes.

When we consider, for example, the excess of a network as a fixed constraint,
a modular organization of the interacting nodes will yield a higher basin
entropy and a shorter average transient time\footnote[39]{A critical relevant
component of $n=10$ nodes and excess $e=10$ has an average entropy of
$\left\langle h\right\rangle \approx0.6$. If two such components with a
combined entropy of $\left\langle h\right\rangle \approx1.2$ merge into a
$n=20$ node component of excess $e=20$ and are randomly rewired, the entropy
decreases to an average value of $\left\langle h\right\rangle \approx0.7$. The
average transient time, on the other hand, increases from $\chi\approx5.5$ to
$\chi\approx12$.}. One might ask if a maximization of the basin entropy or a
minimization of the transient time are desirable features in biological
networks. A short transient time might, for instance, speed up the process of
settling down to a certain cellular state (corresponding to an attractor)
whereas a high entropy would minimize the required number of genes to perform
a classification/decision task. Thus, the ability to rapidly respond to
environmental cues by switching to a particular functional cellular state as
well as the economy with which cellular information processing and
decision-making can be carried out may be evolutionarily selected features of
biological networks.

With the declining costs of microarray and other high-throughput technologies,
time series data will be increasingly available from biological experiments,
enabling network parameters such as basin entropy and transient time to be
studied in a biological context. Our approach to estimate entropy from time
series data sketches one possible way how that might be accomplished.

\begin{acknowledgement}
P.K. would like to thank G.\&M. Krawitz for their constant support and MacT
for helpful instructions. The authors are grateful to S. A. Ramsey and E.
Schweighofer for help with the computing cluster. This work was supported by
NIH/NIGMS No. GM070600, No. GM072855, and No. P50-GM076547 and by the Max
Weber-Programm Bayern.
\end{acknowledgement}

\end{document}